\documentclass{scrartcl}
\usepackage{url}
\usepackage{graphicx}
\usepackage{amssymb}
\usepackage{color}

\begin{document}
\title{An Example for BeSpaceD and its Use for Decision Support in Industrial Automation}

\author{
Jan Olaf Blech 
}

\date{RMIT University, Melbourne, Australia}

\maketitle
\begin{abstract}
We describe our formal methods-based spatial reasoning framework
BeSpaceD and its application in decision support for industrial
automation. In particular we are supporting analysis and decisions
based on formal models for industrial plant and mining
operations. BeSpaceD is a framework for deciding geometric and
topological properties of spatio-temporal models. We present an
example and report on our
ongoing experience with applications in different projects around
software and cyber-physical systems engineering.  The example features
abstracted aspects of a production plant model. Using the example we motivate the use of our framework in the context of an existing
software platform supporting monitoring, incident handling and
maintenance of industrial automation facilities in remote locations.
\end{abstract}

\section{Introduction}

In large-scale industrial automation projects changing demands on collaboration between
stakeholders are an important driver for innovation. These demands cover 
areas such as enterprise architecture over distributed sites, the provisioning of
engineering services and software, testing, verification and monitoring and
generally service support. Mining and manufacturing often involve
large supply chains with different stakeholders. 
Efficient information exchange and interpretation is essential for
cost savings, large-scale technology deployment, and business
services. 
Formal-methods and formal models can help to automate and filter some of the
tasks occurring in operation, maintenance and commissioning of
industrial automation facilities. Formal methods can be integrated and used for decision problems
to support collaboration. They can be used to analyse formal models
such as checking required consistency properties. One, can combine formal methods 
with real-time information to decide on consequences and actions in an
operating systems. For the software frameworks involved, an important question is how to
integrate formal models into them, e.g., by means
of a service bus.

Here, we present a framework aiming at facilitating the
exchange and interpretation of spatio-temporal data and knowledge for industrial
automation. In particular, we are looking at our formal-methods based
BeSpaceD framework. BeSpaceD allows the specification of spatio-temporal
models and reasoning about them. We present the integration into
projects. The models and BeSpaceD based reasoning allow for an easier
integration of multiple sites and facilitate collaboration between
different stakeholders in industrial automation projects.
The main new ideas highlighted in this report comprise:
\begin{itemize}
\item 
The description of formal modeling and reasoning challenges around remotely distributed
  industrial facilities and their BeSpaceD-based modeling and reasoning.
\item 
An example used for describing the application of the use of our formal methods based BeSpaceD framework in
  industrial automation projects.
\end{itemize}
Previously, we have described earlier ideas towards BeSpaceD \cite{bespaced0,bespaced1}. Here,
we focus on the industrial automation domain and analyse an existing
solution provided by us.

\subsection*{Overview}
Our BeSpaceD framework is described in
Section~\ref{sec:bespaced}. An example model/case is given in
Section~\ref{sec:example}. An application of BeSpaceD and formal
methods-based reasoning in the collaborative
engineering framework is described in
Section~\ref{sec:colleng}. 
We discuss related work on
Section~\ref{sec:relwork}. Finally, a conclusion is provided in Section~\ref{sec:concl}.

\section{Spatio-Temporal Reasoning using BeSpaceD}
\label{sec:bespaced}

BeSpaceD \cite{bespaced1,bespaced0} is a constraint solving and
non-classical model checking framework. It is organized as a library
and specification language with a focus on spatio-temporal
properties. In the industrial automation context discussed in this report, we use BeSpaceD to specify
industrial plant models and for dynamically
and statically deciding on consequences of an
event / alarm occurring in a system. We semantically -- in the
spatio-temporal context -- interpret (series)
of alarms occurring in the system and for retrieving and processing relevant information.
The BeSpaceD framework comprises
(i) a modeling language focusing
on space and time, and 
(ii) a library to reason about models and their properties. Library
functions comprise, spatio-temporal decision procedures such as intersections, state-space
exploration, abstraction and reduction. These are combined for checking properties of the models expressed in the BeSpaceD
modeling language and for deciding on actions and consequences.
Since BeSpaceD based constraint solving can be done using and combining
library functions, we are flexible in writing customized checking procedures.
Our BeSpaceD modeling language allows the time or automata based
behavioral description of entities. Description integrate spatial
(coordinates or topological) characteristics. Typical descriptions include
 availability areas and schedules, capabilities, events and states.

BeSpaceD-based checking is done by using a series of steps where
BeSpaceD functions and
language elements are combined. These steps comprise preprocessing,
abstraction and derivation of verification conditions. Verification
conditions are checked by using tools like SAT~\footnote{we have
  implemented a connection to  Sat4j:
  \url{http://www.sat4j.org/}} and SMT solvers (a connection to
z3~\cite{z3} exists) and specialized algorithms. The
creation of verification conditions
requires the encoding of spatio-temporal properties into SMT-like
formulas, e.g., by specifying large conjunction of predicates each one
indicating a spatio-temporal coordinate: (x,y,z,time) at a predefined
resolution. Furthermore, we have integrated to notion of ownership and
over and underapproximation for reasoning about safety into these
predicates \cite{bespaced1}.

A small excerpt of our BeSpaceD language definition is shown in
Figure~\ref{fig:scaladef} to give a look and feel of the principal
specification idea. Constructors for abstract datatypes in
Scala can be combined to create a model. The excerpt shows
constructors for logical operations, timepoints and intervals, events,
probabilities, geometric elements, and topological elements.
\begin{figure*}
{\footnotesize
\begin{verbatim}
case class OR (t1 : Invariant, t2 : Invariant) extends Invariant
case class AND (t1 : Invariant, t2 : Invariant)  extends Invariant
case class NOT (t : Invariant) extends Invariant
case class IMPLIES (t1 : Invariant, t2 : Invariant)  extends Invariant
...
case class TimePoint [T] (timepoint : T) extends ATOM 
case class TimeInterval [T](timepoint1 : T, timepoint2 : T) extends ATOM
...
case class Event[E] (event : E) extends ATOM
case class Owner[O] (owner : O) extends ATOM
case class Prob (probability : Double) extends ATOM
case class ComponentState[S] (state : S) extends ATOM
...
case class OccupyPoint (x:Int, y:Int) extends ATOM
case class OccupyBox (x1 : Int,y1 : Int,x2 : Int,y2 : Int) extends ATOM
case class OccupyCircle (x1 : Int, y1 : Int, radius : Int) extends ATOM 
...
case class Edge[N] (source : N, target : N) extends ATOM 
case class Transition[N,E] (source : N, even : E, target : N) extends ATOM 
...
case class TRUE() extends ATOM
case class FALSE() extends ATOM
\end{verbatim}
}
\caption{Abstract datatypes for BeSpaceD}
\label{fig:scaladef}
\end{figure*}
Different levels of abstraction can be distinguished: for example,
geometric boxes can be broken down into sets of geometric points.
BeSpaceD models may be manually written directly using the BeSpaceD modeling language.
Alternatively, generation of behavioral models from code is possible. Here, BeSpaceD descriptions are created
  by executing customized code pieces. 

BeSpaceD comprises a variety of functionality. Important for
this report are:
\begin{itemize} 
\item
Abstraction functionality comprises (i) the aggregations
  of information for time points into time intervals which are safe-overapproximations, (ii)
  operations that support the safe over- and underapproximation of
  geometric objects. 
\item Verification goal generation supports (i) the breakdown of
  geometric objects associated with time and space into predicates
  characterizing time and space points and point sets, (ii) the
  generation of input for SAT (individual point predicates) and SMT
  (point sets) solvers and other specialized algorithms.
\item Solving verification goals supports operations on
  spatio-temporal objects such as
  inclusion and
  intersection.
\item Management of objects comprises assigning ownerships to
  spatio-temporal regions, topological objects and other
  structures. This also allows the management of aspects that require
  safe abstractions with respect to over- and
  underapproximation. Furthermore, various search and model
  restructure operations are supported.
\end{itemize}
Safe overapproximations for time and geometric space are useful, for
guaranteeing the absence of collisions, i.e., the model suggests
greater expansion of an object than its real physical properties. On
the other hand, for ranges, we may use a safe
underapproximation. Both, under and over-approximations for geometric and
topological information can be kept in the same model and are distinguished by the
ownership predicate from Figure~\ref{fig:scaladef}.  

\section{Example Case: Remote Robot Interactions}
\label{sec:example}
\begin{figure}
 \centering\includegraphics[width=0.4\textwidth]{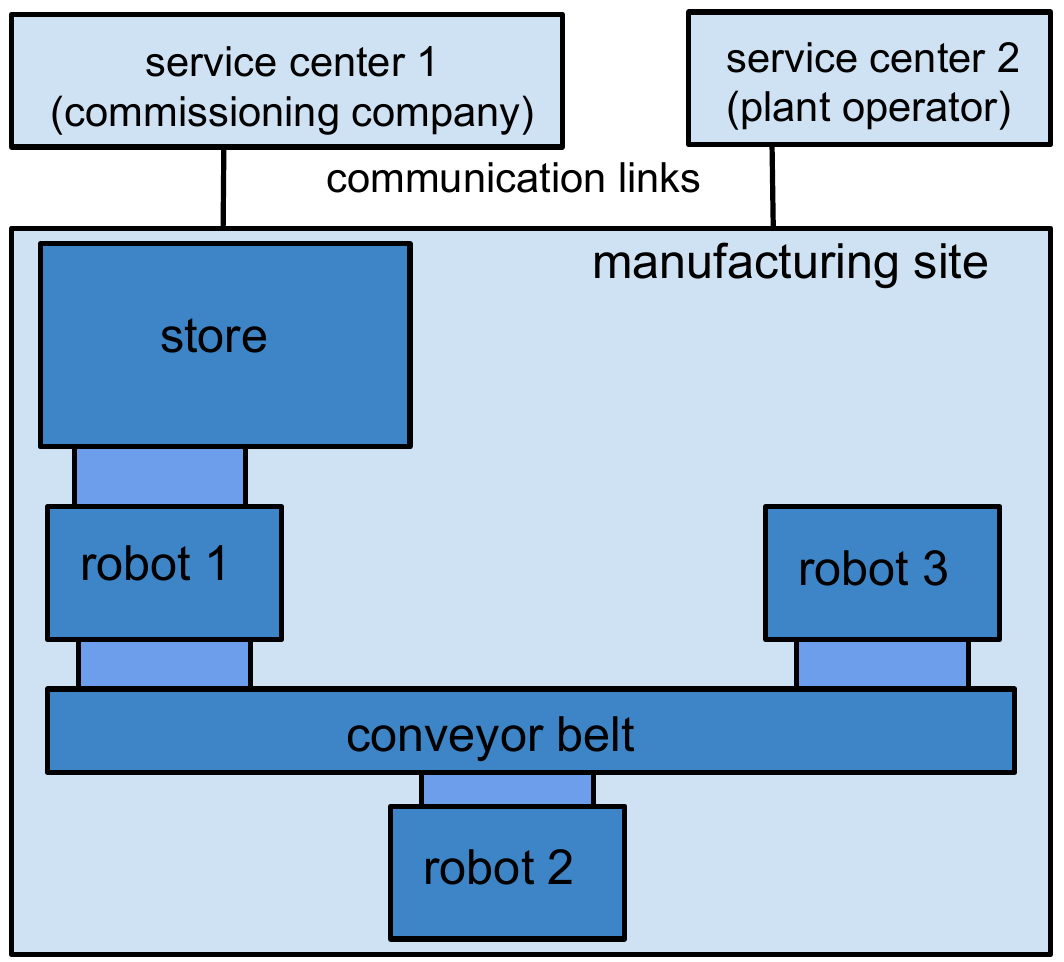} ~~~~~
\includegraphics[width=0.4\textwidth]{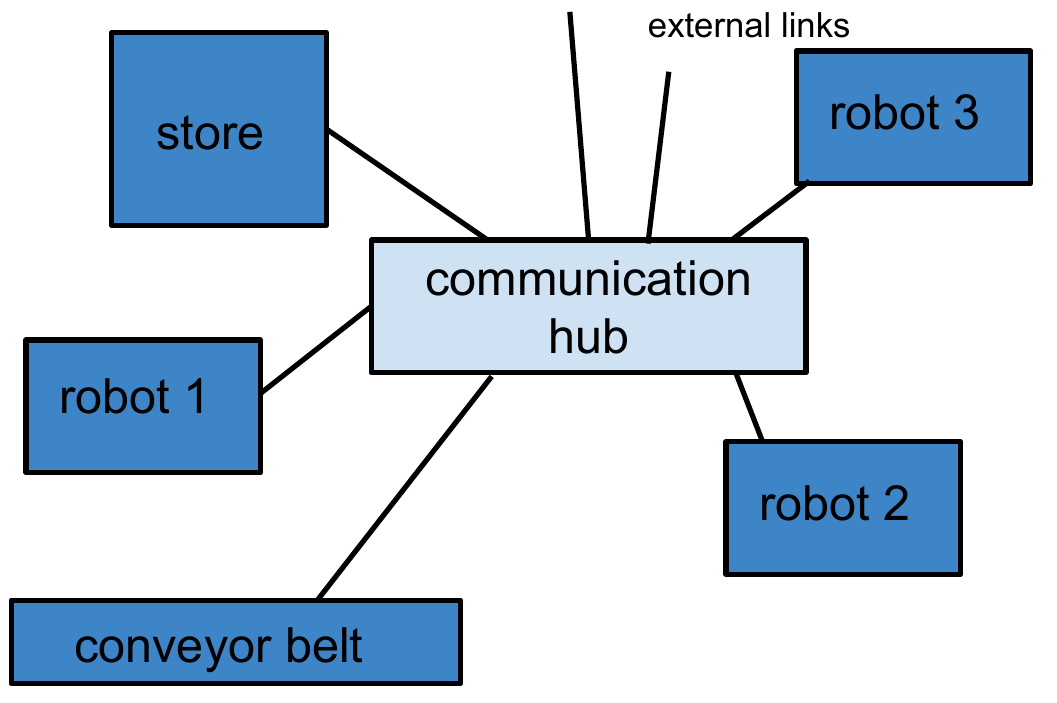} 
\caption{Physical influences overview and communication topology for the manufacturing site}
\label{fig:ex1}
\end{figure}

\begin{figure}[t]
\centering\includegraphics[width=.99\textwidth]{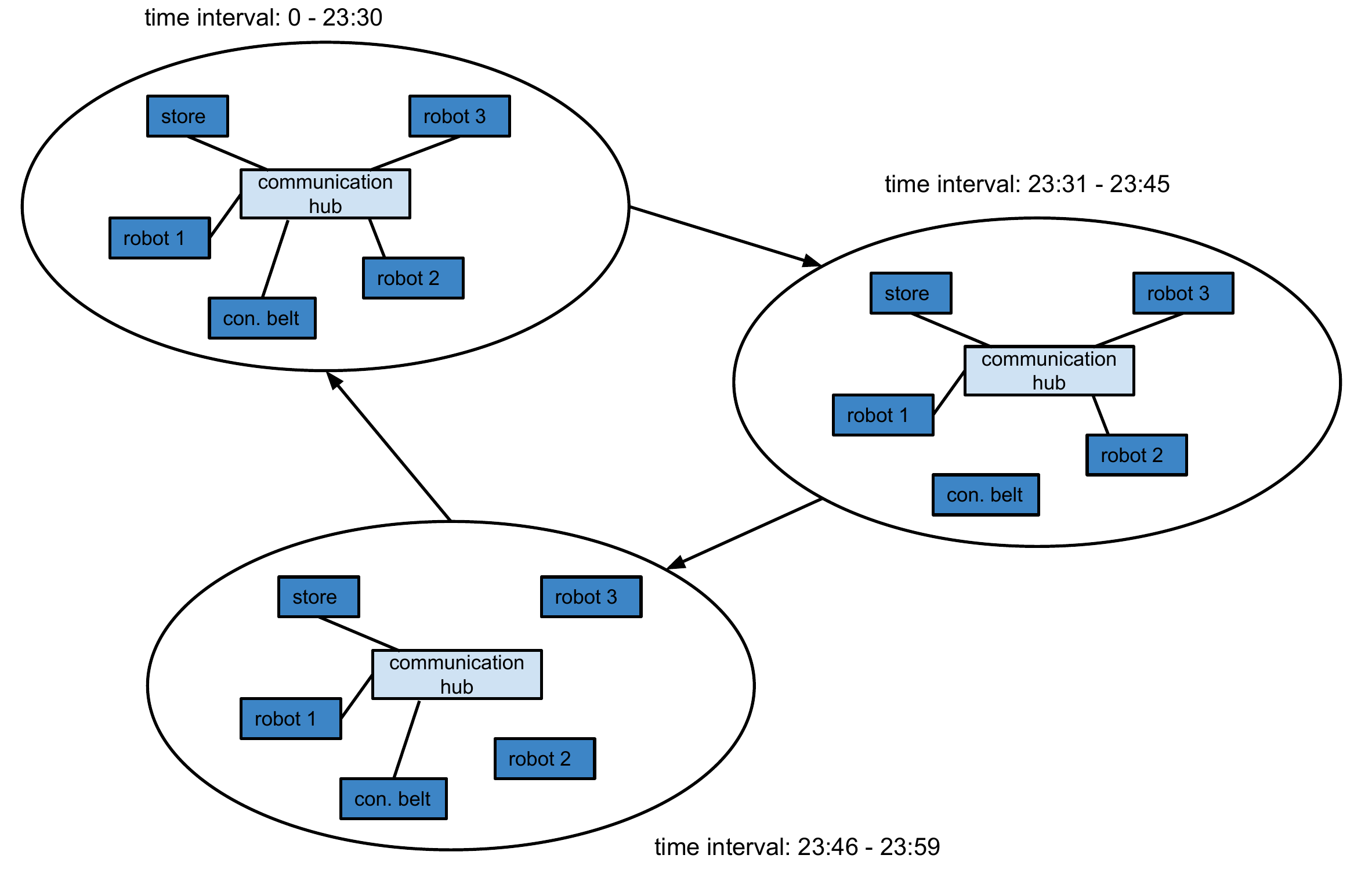}  
\caption{Communication model for the manufacturing site}
\label{fig:ex3}
\end{figure}

We present an example scenario in this section to give a better understanding of our  industrial automation use-cases.
We describe a static model, dynamic aspects, example properties and their verification.
Robots are deployed in a remote processing plant. The plant can be
observed and reprogrammed via a remote service center. An overview on
the physical setup is given in the left part of Figure~\ref{fig:ex1}.
Three robots (robot 1, robot 2, robot 3) can be distinguished operating on a conveyor belt. Together with a store, they are forming the manufacturing site. The manufacturing site has communication links to two service centers: one for the company who commissioned the installation and one for the company operating the plant.
The communication links within the manufacturing site follows a
star-topology which can be described as a graph seen in the right part
of Figure~\ref{fig:ex1}.
The communication hub does not appear as a physical component in the
left part of Figure~\ref{fig:ex1}. Furthermore, communication between some robots, the
conveyor belt and the communication hub is shut down in regular
intervals for maintenance reasons. We include this as temporal aspects
in our communication model for the manufacturing site as shown in Figure~\ref{fig:ex3}.
Even in this part of the example, different levels of modeling have different characteristics:
\begin{enumerate}
\item
No physical interaction is possible between different sites, physical
interactions at the site level is possible, but undefined (may or may
not occur) at this modeling level. This is formally modeled using an
empty graph. The fact that between different sites, communication is possible along the
communication lines is modeled using a graph with three nodes (one for each site: service center 1, service center 2, manufacturing site) and two edges: 
(service center 1, manufacturing site), 
(service center 2, manufacturing site).
\item
Another layer describes possible interactions between: robot 1, robot 2, conveyor belt, storage. Each one is a node in the graph. Communication is not shown in the Figure 1, but in Figures 2 and 3. For the physical influence graph, we have: 
(robot 1, conveyor belt), 
(robot 2, conveyor belt), 
(robot 3, conveyor belt), 
(robot 1, store).
No physical interaction is possible in the absence of an edge. The physical influence graph indicates possible influences.
\end{enumerate}
The communication (time and local communication included) from Figure~\ref{fig:ex3} is formalized in BeSpaceD as
shown in Figure~\ref{fig:ex4}.
\begin{figure}
{\footnotesize
\begin{verbatim}
def midlevelcommlinkgraph = IMPLIES(Owner("midlevelcommgraph"),BIGAND(
     IMPLIES(TimeInterval(TStandardGMTDay(00,00,00),TStandardGMTDay(23,30,59)),
       BIGAND(Edge("ComHub","Robot1")::Edge("ComHub","Robot2")::
         Edge("ComHub","Robot3")::Edge("ComHub","Store")::Edge("ComHub","ConvBelt")::
             Nil))::
     IMPLIES(TimeInterval(TStandardGMTDay(23,31,00),TStandardGMTDay(23,45,59)),
       BIGAND(Edge("ComHub","Robot1")::Edge("ComHub","Robot2")::
         Edge("ComHub","Robot3")::Edge("ComHub","Store")::Nil))::
     IMPLIES(TimeInterval(TStandardGMTDay(23,46,00),TStandardGMTDay(23,59,59)),
       BIGAND(
        Edge("ComHub","Robot1")::Edge("ComHub","Store")::Edge("ComHub","ConvBelt")::
            Nil))::Nil))
\end{verbatim}
}
\caption{Communication graph in BeSpaceD}
\label{fig:ex4}
\end{figure}
An additional modeling layer is available encapsulating
fire and motion detection sensors and their ranges. We assume, that
fire and motion detection sensors are deployed in a grid like fashion
in the factory hall. The communications of this sensor network is done
using wireless technology and thus does not depend on a physical
link.
Each sensor is annotated with a detection range indicated by the circle.
We also have geometric models, for fine grained interactions between different entities. Figure~\ref{fig:ex3b} shows an interaction sequence of robot 2 handling and modifying a work piece on the conveyor belt. The figure shows on overapproximation of occupied space for each step. This model encapsulates spatial impact of an action sequence performed by robot 2 on arrival of a work piece. Actual arrival times of work pieces are not part of the static model and are dynamic information. 
 \begin{figure}[t]
\centering\includegraphics[width=.75\textwidth]{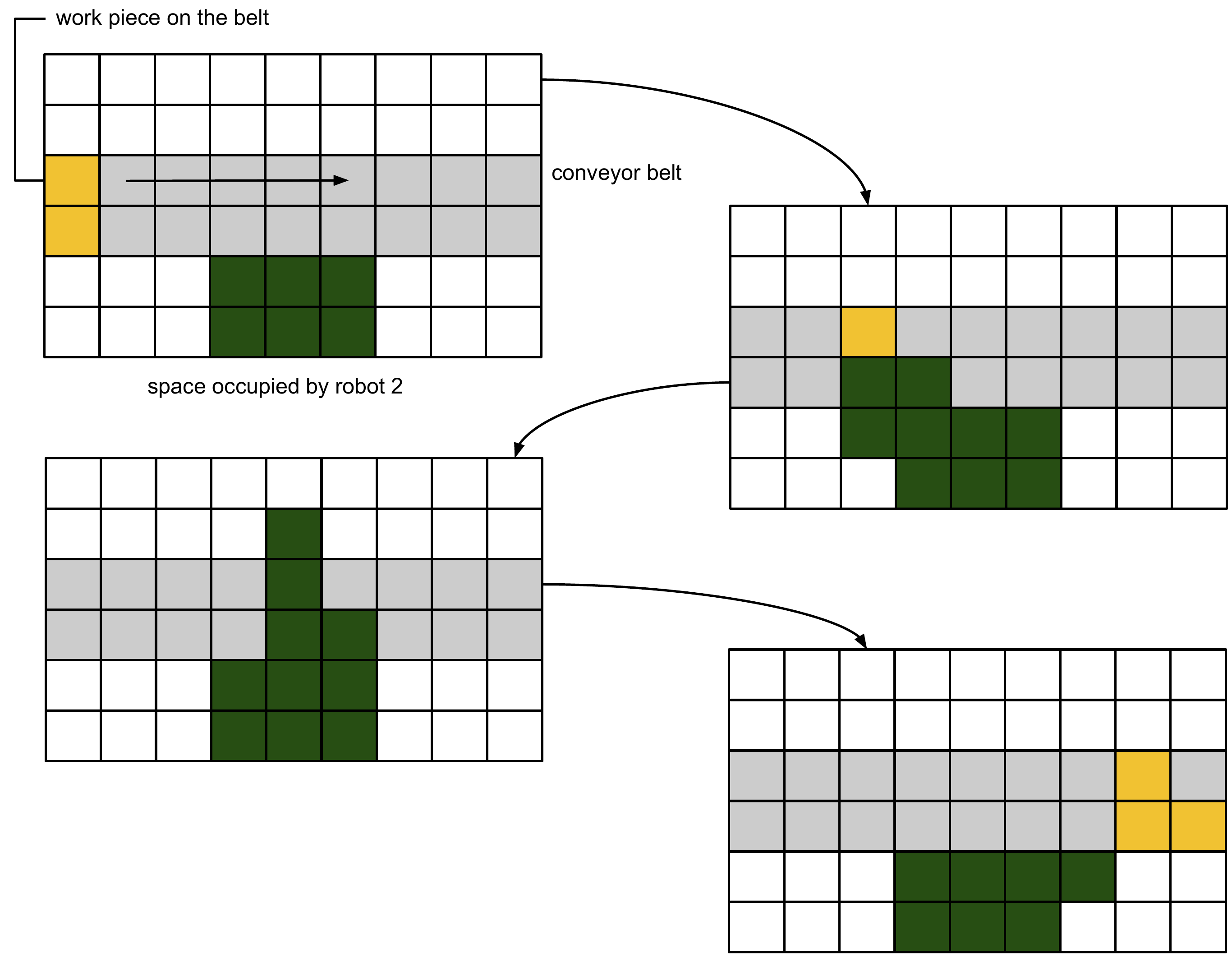} 
\caption{2D Geometric abstraction of robot 2 interacting with a work
  piece}
\label{fig:ex3b}
\end{figure} 
Furthermore, a Scala code fragment generating the corresponding
BeSpaceD model with boxes that overapproximate the spatial occupation for the movement is
shown in Figure~\ref{fig:ex3d}. The code makes use of the actual
positioning functions of the objects: {\tt moveRobot2} (not shown) and the
linear conveyor belt movement {\tt moveWorkPiece}.
\begin{figure}
{\footnotesize
\begin{verbatim}
def moveWorkPiece (time : Int) : (Int,Int,Int,Int) ={
  if (time < 1000 && time > 0) {
     return(moveObjOnConvBelt(time),100,moveObjOnConvBelt(time)+20,120)} 
  return (0,0,0,0)}

def mR2bespaced[E] (e: E, t: Int, a : Int, b : Int, c: Int, d: Int) : 
Invariant ={return (IMPLIES(TimeStamp(TERTP(e,t)),OccupyBox(a,b,c,d))) }
		
def createTrajectoryAbstraction1() : Invariant ={
  var retinv1 : List[Invariant] = Nil
  var retinv2 : List[Invariant] = Nil
  for (i :Int <-  0 to 100) {
     retinv1 ::= (moveRobot2(i) match {case (a,b,c,d) => 
         mR2bespaced("ConvAct",i,a,b,c,d)})
     retinv2 ::= (moveWorkPiece(i) match {case (a,b,c,d) => 
         mR2bespaced("ConvAct",i,a,b,c,d)})}
  return(BIGAND(IMPLIES(Owner("Robot2_Space"),BIGAND (retinv1))::
          IMPLIES(Owner("WorkPiece_Space"),BIGAND (retinv2))::Nil)) }
\end{verbatim}
}
\caption{Scala code generating BeSpaceD models workpiece / robot 2
  interaction (extract)}
\label{fig:ex3d}
\end{figure}
Different levels of modeling space are shown:
(i) The course topological classification of sites;
(ii) the more fine-grained geometric formalization of the remote
processing plant;
(iii) the even more fine-grained modeling of the space around the robot.
An arbitrary number of different aspects can be distinguished for each level of modeling space. In the example we have the following aspects:
(i) physical interaction,
(ii) communication (in the example, two disjunct aspects are provided)
(iii) detection ranges of sensors.

\section{Decision Support for Collaborative Engineering and Related Use-Cases}
\label{sec:colleng}

In this section we describe the highlights of formal-methods based
application of BeSpaceD to the collaborative engineering project \cite{etfa,etfa2015}.
Our collaborative engineering project focuses on enabling the
exchange of data and knowledge for remote plant operation, services
and maintenance. At its core, we use a BeSpaceD based decision support framework, that
provides relevant information to plant operators, engineers, other
staff and stakeholders. It is focused around the handling of events
and providing appropriate information. In
industrial automation events comprise {\it alarms}. These are typically issued by a control
system such as a SCADA (supervisory control and data acquisition)
system. For example, events can be based on  sensor values deviating from a pre-defined
range or manual
triggering. Our BeSpaceD based decision support relies on formal models encapsulating the
semantics of system components and related ontologies.

\subsection{Framework Architecture Overview}
\begin{figure*}[t]
\centering
\includegraphics[width=.71\textwidth]{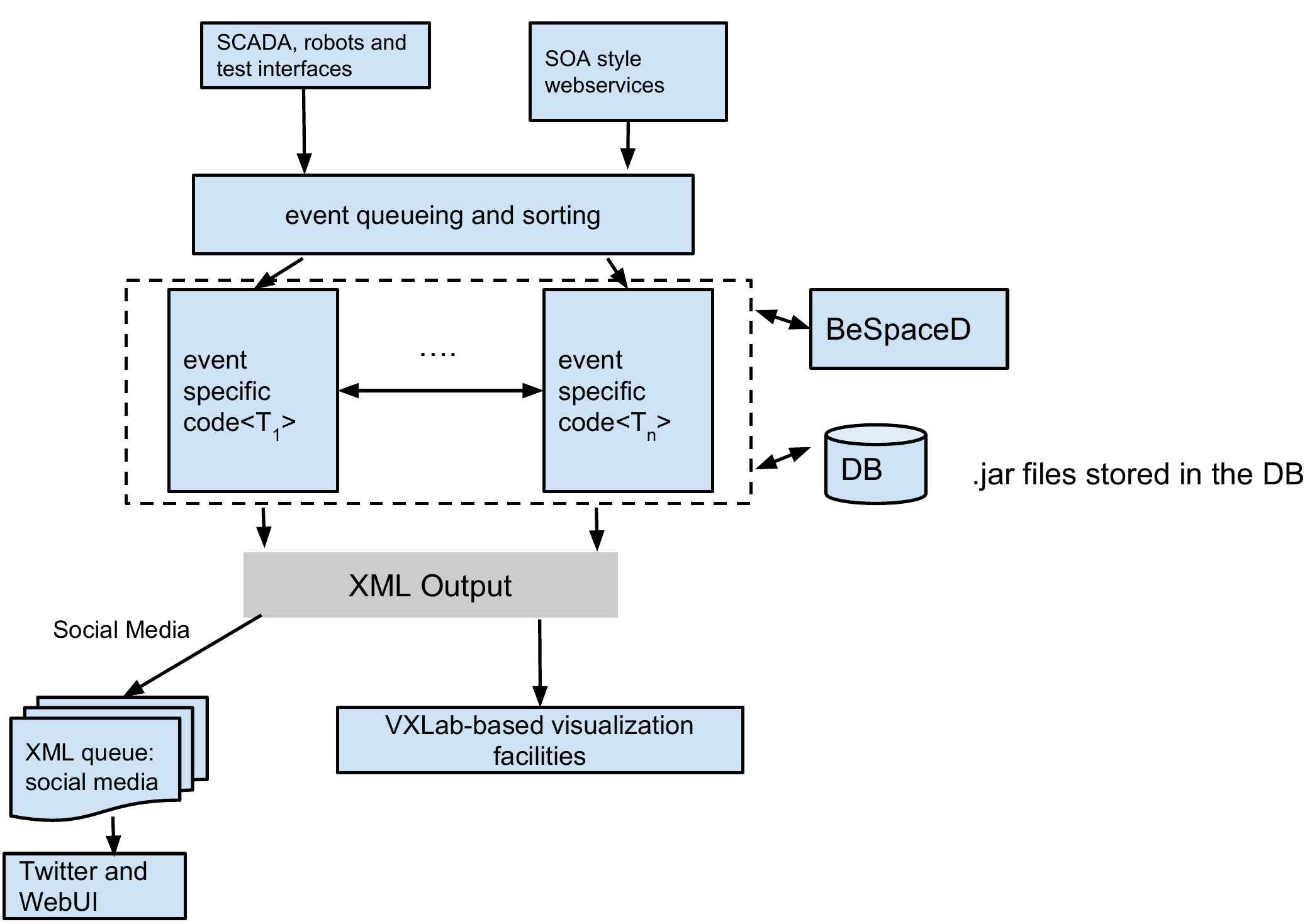} 
\caption{Collaboration service bus, following \cite{etfa2015}, with
  additional  extensions}
\label{fig:insidemt}
\end{figure*}

Figure~\ref{fig:insidemt} gives on overview on the implemented 
collaboration support framework: from event generating devices to the
display of information to stakeholders:
\begin{enumerate}
\item
Events are collected from different devices, such as SCADA systems,
robots and from webservices
\item
Events are preprocessed, queued and sorted.
\item
The event specific handling is
parallelised. Based on spatio-temporal models, we derive appropriate actions. The event specific code is also
 emitting the XML code for visualization.
\end{enumerate}
Our  event specific handlers
comprise BeSpaceD based reasoning. Using models provided as .jar files
and dynamic information from the events, we generate queries inside
the event specific code to decide questions as described in the
previous sections. Based on this, appropriate
information is selected for experts, collaborators, and other
stakeholders.

\begin{figure*}
\centering
\includegraphics[width=0.59\textwidth]{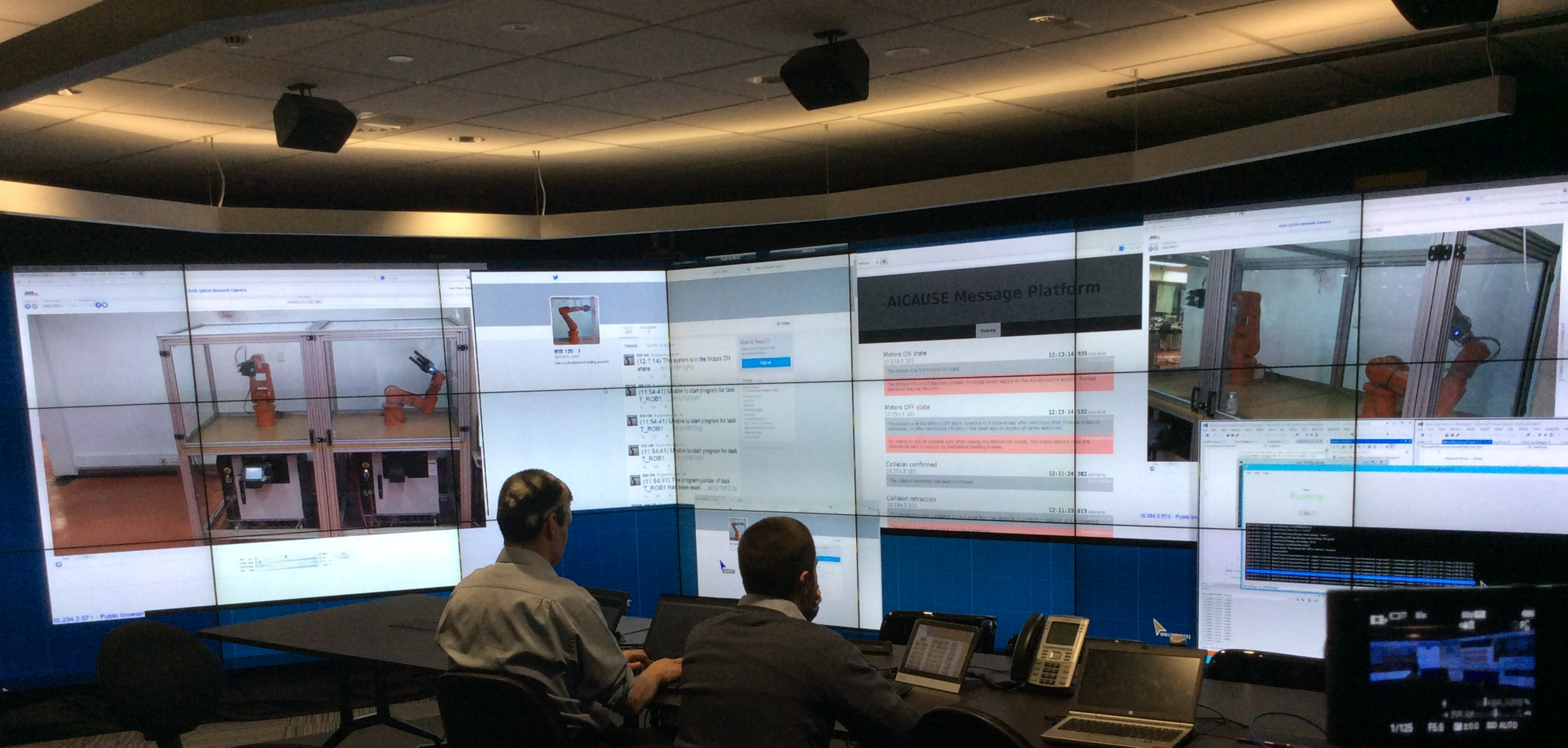} 
\caption{Our multi-screen visualization \cite{enase}}
\label{fig:vxlab}
\end{figure*}

Internally, event specific handlers share a global state. This state is used to
share information between event specific code and for tracking the event history.
The display of information is triggered by emitting XML code. The XML
code is interpreted by a visualization manager for display of selected
information in device specific ways. This can comprise mobile devices, workstations or our large screen visualization facilities (cf. Figure~\ref{fig:vxlab}).

\subsection{Example Use-Case}
\label{sec:exuc}

Our example use-case builds on the modeling techniques and the example
model used in Section~\ref{sec:example}.
To illustrate our framework in an event handling action, we are
providing the following example scenario. The presented use-case is
not part of the collaborative engineering project and is only used to
exemplify the use of formal-methods in the framework, see
also \cite{etfa,etfa2015} for a different use-case described from an
industrial engineering point of view in remote
surveillance  that is a part of
the collaborative engineering project. 
We describe our framework responding to an
event, an alarm triggered by a malfunction such as a failure of a
robot or a communication device provided in Section~\ref{sec:example}.
Our goal is to provide relevant information to staff,
stakeholders, experts,
and/or engineers. Information needs to be provided in a concise
way. We can face situations, where many alarms arrive
in a short time. Information for display has to be filtered so that
humans are not overburdened. 
\begin{enumerate}
\item
A sensor provides data indicating a malfunction of robot 2 in our example plant. We check the confidence by
investigating historical data. 
\item
If there is enough confidence, we generate an alarm this is provided
as an event to our framework. 
\item
Using our BeSpaceD based models of the plant and the reasoning
functionality, we have implemented functions that find nearby machines
and devices through the geometric and topological models. Furthermore,
we can use this information to decide whether additional actions need
to be taken. For example, our collaboration platform can automatically
match experts to the situation and offer resource conflict
resolution. Our semantic models also provide
information on states of machinery. This may be used to identify
additional possible dangers, possible interactions, physical
locations, and possible effects on the surrounding area.
In addition to our semantic models, the event specific code
can take information from databases  and rely on
real-time information from streaming sources. 
\item
In the next step, we
select incident relevant information for display to humans. We
use an XML-based language to encode commands for triggering changes to
display information on mobile, devices, normal workstations and large
scale visualization screens.
\end{enumerate}
In collaborative engineering, the information displayed comprises profiles and other data
stored in SharePoint as well as maps. The SharePoint-based data is displayed in browser
windows managed by our framework. 

\subsection{Evaluation}

In addition to the industrial automation decision support usages
discussed in the last section, we have applied BeSpaceD in combination
with other model checkers to a
number of different projects. In \cite{fesca2014} and \cite{apscc}, we
are also dealing with models that comprise a large amount of time and
space based specifications for industrial automation. The focus is on verification of
consistency conditions at design time of a system to avoid possible
collisions and injuries using probabilistic and
non-probabilistic models. 
In \cite{han2015} we have applied BeSpaceD for assuring correct
sensor ranges in factory hall scenarios. Here, some dynamic aspects
are integrated.
Different formal methods exist for supporting collaboration. In our
case, we provide distinct functionality to support spatio-temporal
decisions. In our case, we can map most problems to SAT / SMT
and geometric inclusion problems. These can be solved by applying the
appropriate tools or by state-space exploration.
A library style framework as opposed to a tool in a tool chain allows
the flexible combination of functionality. In our case studies we
found the following cases:
\begin{itemize}
\item
Dividing a model into smaller units and keeping them in databases
increases the lookup and update speed and thereby the decision speed.
\item 
Combination of different functionality can be realized for all
  kinds of event specific code in the collaborative engineering
  project, which allows for adding new events and devices that were not
  present at the design time of the system.
\end{itemize}
BeSpaceD models can either be hand-written or generated out of code
(see e.g., \cite{fesca2014} for an application). Hand-writing is
  can be regarded as  working with a domain specific language and
  requires some expertise.

\section{Related Work}
\label{sec:relwork}

BeSpaceD enables spatio-temporal reasoning. Existing specification and
reasoning techniques comprise process algebra like
formalisms \cite{cardelli03,cardelli04} and \cite{haar}. A type system in connection
with this work has been introduced in \cite{cairestypes}. Applications
comprise concurrency
and ressource control. Another framework for describing
hybrid programs with stochastic features is described in \cite{platzer2011}.
A verification tool to check
properties based on this formalism is described \cite{slmc}. In our work, we are
more focused on a domain specific solution for industrial automation.
We are more restrictive by concentrating on spatio-temporal properties
with respect to geometry and topology and focus on tailoring our
formalism and related decision techniques
for industrial automation. 
Highly specialised solutions for reasoning about geometric constraints
are important in robot path planning. This has been studied for
decades, e.g., \cite{robots1,robots2}.  In addition different kinds of spatial logics and means to
reason about them have been studied (e.g., \cite{hirschkoff,Bennett})
including work on decidability (e.g., see decidability results
in \cite{zilio}). Complementing the time and geometry focus on the reasoning
side of our framework, a strong focus on topological models has advantages in areas such as security analysis \cite{topology}. 
Spatial types as classification elements for managing
geometric objects are also important in databases \cite{spatialtype1} and in Geographic Information
Systems \cite{spacegis}.

\section{Conclusion and Future Work}
\label{sec:concl}

We have shown the application of our formal methods-based spatial
reasoning framework BeSpaceD in industrial automation. BeSpaceD works
on spatio-temporal models. Analysis results are used in decision
making. 
Future work comprises more expressive modeling and shifting more
functionality from use-cases into the BeSpaceD library thereby
generalizing the framework. Ongoing work comprises various work around
the analysis and checking of industrial automation
models (e.g., \cite{cyberbeht}). Furthermore, a connection to our work on the specification of
PLC software for controlling machinery through behavioral types
\cite{wenger} and complete semantical specifications \cite{borja,borja2} is ongoing.

\bibliographystyle{eptcs}

\end{document}